\documentstyle[aps,prb,multicol,epsf]{revtex}  
\begin{document}
\title{``Scars'' in parametrically excited surface waves} 
\author{Oded Agam$^a$ and Boris L. Altshuler$^b$}
\address{$^a$ The Racah Institute of Physics, The Hebrew University,
Jerusalem 91904, Israel. \\ 
$^b$ Physics Department, Princeton University, Princeton, NJ 08544 \\
and  NEC Research Institute, 4 Independence Way, Princeton, NJ 08540.
}
\maketitle
\begin{abstract}
We consider the Faraday surface waves of a fluid 
in a container with a non-integrable
boundary shape. We show that, at sufficiently low frequencies, 
the wave patterns are ``scars''  selected by 
the instability of the corresponding periodic orbits,  
the dissipation at the container side walls, and interaction effects
which reflect the nonlinear nature of the Faraday waves.
The results explain the observation of a limited number of scars
with anomalous strengths in recent experiments by Kudrolli, Abraham 
and Gollub.
\end{abstract}
\begin{multicols}{2}
A prominent feature of 
wave chaos is the ``scar'' phenomenon\cite{Heller84}. 
Namely, wave functions are concentrated along the 
short periodic orbits of the underlying 
classical system. In general, the phenomenon appears whenever the 
Eikonal approximation holds, and the corresponding
rays exhibit ballistic chaotic dynamics. Since their 
discovery by numerical studies\cite{Heller84}, scars have been observed 
experimentally in microwave cavities\cite{microwaves}, 
tunneling diodes\cite{diodes}, and capillary waves\cite{faraday}. 

However, theoretical studies of 
scars\cite{PreviousTheory} have not addressed several 
factors which may be present in real systems, such as dissipation,
external forcing, nonequilibrium effects, and 
interactions associated with nonlinear terms of the wave equations.
In this Letter we study a system
where these factors are significant: the surface
waves\cite{review} in a container of non-integrable shape.

This work was motivated by recent experiments of Kudrolli 
et al.~who studied Faraday waves in a stadium 
shaped container\cite{exp}. The surface waves were
excited parametrically by oscillating the container in direction
perpendicular to the fluid surface at rest. The experimental findings
are quite intriguing, see Fig.~1: I.
Out of all known scars of the stadium billiard\cite{Heller84}
only three scarring
patterns were identified; II. These patterns appear in, 
approximately, 90$\%$ of the cases, and their magnitudes, compared 
to the random background, is
unexpectedly large; III. While probability densities associated with
eigenstates of the stadium billiard were always symmetric, 
nonsymmetric wave patterns have been also observed.

The purpose of this Letter is to explain this behavior. In particular, 
we show that the excitation threshold for observation of
a scarring pattern is governed by the Lyapunov exponent of
the corresponding periodic orbit, bulk dissipation, and 
dissipation at the boundary of the vessel. These factors 
suppress most of the scarring patterns
as well as the random background. The amplitudes of those scars which
still can be excited are determined by nonlinear effects. 
These effects (interactions) also break the symmetry 
between degenerate scars, which results in
nonsymmetric wave patterns.

To begin with, consider the surface waves of an infinite system 
with an infinite fluid depth. Within the linear approximation, 
the amplitude of a plain wave, $a_{\bf k} e^{i {\bf k \cdot r}}$, 

{\narrowtext    
\begin{figure}      
  \begin{center}     
\leavevmode     
        \epsfxsize=8.0 cm        
         \epsfbox{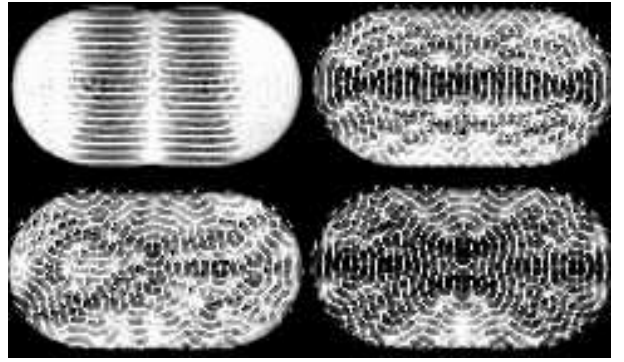}       
\end{center}    

	\caption{Surface wave patterns in a stadium shaped
container at various external frequencies.
Adapted from Ref.~7}               
\end{figure}        
}
\noindent
satisfies the equation\cite{review}:
\begin{eqnarray}
\frac{\partial^2 a_{\bf k}}{\partial t^2} 
+ 2 \gamma_{b}
\frac{\partial a_{\bf k}}{\partial t} +
 \omega^2_{k} a_{\bf k}=0, \label{amplitude}
\end{eqnarray}
where $k$ is the wave number, $\gamma_{b}= \nu k^2$ is the bulk
dissipation rate, and $\nu$ is the viscosity. The angular frequency, 
$\omega_k$, satisfies the dispersion relation:
\begin{eqnarray}
\omega_k^2= g k + \frac{\Sigma}{\rho} k^3, \label{disperssion}
\end{eqnarray}
where $g$ is the acceleration of gravity, $\Sigma$ is the surface 
tension, and $\rho$ is the fluid density. 

In a finite system, the boundary causes two
effects. First, it leads to quantization of the wave 
vector $k \to k_\alpha$ (thus also $\omega_k \to \omega_\alpha$). 
When the container is rim full, the fluid surface 
is believed to be pinned to boundary\cite{Benjamin79}.  Thus the
eigenmodes of the surface waves, $\psi_\alpha({\bf r})$, 
are solutions of the Helmholtz equation in a domain of the 
container shape with Dirichlet boundary conditions.
 
The second effect of the boundary is an  
additional dissipation due to the viscous flow 
in the vicinity of the container side walls. This dissipation,
in a system with size $L$, is of
the order $\sqrt{\nu \omega}/L$.
For large containers it is negligible 
compared to the bulk dissipation rate. However, 
in small systems and for specific wave patterns, this dissipation becomes
 comparable to $\gamma_b$.

Parametric excitation of surface waves, due to vertical 
shaking, is accounted by introducing time dependent acceleration,
$g \to g+ A \cos (  2\omega t)$, into Eq.~(\ref{amplitude}).
The parameter space $(\omega, A)$ of the resulting Mathieu equation 
is characterized by tongues of instability where $a_{\bf k}(t)$ grows
exponentially in time. This growth saturates due to  nonlinear effects.

To take these interactions into account,
it is convenient to employ a formalism analogous to
second quantization\cite{LvovBook}. 
For simplicity, we first consider an infinite system. 
Let us introduce the amplitude variables,  $a^*_{\bf k}$ and $a_{\bf k}$, 
analogous to boson creation and annihilation 
operators at  momentum state ${\bf k}$.
These variables are $c$-numbers which can be viewed 
as boson operators in the limit of a large number 
of particles. The equations of motion, then, take the form: 
\begin{equation}
\frac{\partial}{\partial t} a_{\bf k} + \gamma_b a_{\bf k}
= - i \frac{\partial {\cal H}}{\partial  a^*_{\bf k}}, \label{eqm} 
\end{equation}
where ${\cal H}$ is the Hamiltonian of the system which consists of
three terms: ${\cal H}={\cal H}_0 +
{\cal H}_P + {\cal H}_{\mbox{int}}$. The first component,
${\cal H}_0 = \sum_{\bf k} \omega_k  a^*_{\bf k} a_{\bf k}$,
describes the unforced surface waves
in the linear approximation, i.e.~Eq.~(\ref{amplitude}).

The second component, ${\cal H}_P$, is the pumping Hamiltonian 
responsible for 
the parametric excitation of the Faraday waves. Confining our
attention to excitations near the threshold of the first
instability tongue of the Mathieu equation, we can write
 ${\cal H}_P$ as
\begin{equation}
{\cal H}_P = \frac{1}{2}\sum_{k} h 
a^*_{\bf k} a^*_{-{\bf k}} e^{i 2\omega t} + C.c.; ~~~ 
h =  \frac{A k \omega^2}{4 \omega_k},
\end{equation} 
where $A$ and $2 \omega$ are the amplitude and the frequency of the
vibrations.
In the quadratic approximation, ${\cal H}\simeq
{\cal H}_0 +{\cal H}_P$, the equations of motion (\ref{eqm}) yield
\begin{eqnarray}
a_{\bf k}, a_{\bf k}^* \propto e^{\sigma t},~\mbox{where}~~
\sigma = -\gamma_b + \sqrt{h^2- (\omega-\omega_k)^2}. 
\end{eqnarray}
Thus 
the threshold for instability, at resonance $(\omega=\omega_k)$, is 
$h\geq \gamma_b$. Under this condition, the rate of
energy pumped into the system exceeds the dissipation rate. 

The third component of the Hamiltonian, ${\cal H}_{\mbox{int}}$, 
is an interaction term. In the mean field approximation it 
takes the form\cite{LvovBook}
\begin{eqnarray}
{\cal H}_{\mbox{int}}= \sum_{\bf k k'}
I^{(1)}_k(\theta)
|a_{\bf k}|^2 |a_{\bf k'}|^2 + I^{(2)}_k(\theta)
a_{\bf k'}^* a_{- \bf{k'}}^* a_{\bf k} a_{-{\bf k}}, 
\end{eqnarray}
where $I^{(j)}_k(\theta)=
T^{(j)}_k(\theta)+ i \Gamma^{(j)}_k(\theta)$ are complex
matrix elements which depend on the length of the vectors, 
$k= |{\bf k}|= |{\bf k'}|$ as well as the angle, $\theta$, 
between them. The exact form of these matrix elements has 
been derived by Milner\cite{Milner91}. Here we only note
that the interaction is local and $T$ is of order 
$\omega k^2$, while $\Gamma \propto \nu k^4$.

Several mechanisms limit the exponential growth of
the wave amplitude. They are related to different 
components of the interaction matrix elements. For instance,
$ \Gamma^{(j)}_k(\theta)$ are associated with
nonlinear damping\cite{Milner91}, while $T^{(2)}_k(\theta)$ governs the
missphasing mechanism\cite{Zakhsrov75}. 
Missphasing is the phase difference between the pump and the excited 
wave, which appears due to the nonlinearity and suppresses the energy 
absorption.

The actual limiting mechanism is determined by the system size
and the excitation frequency. In large systems, boundary effects
can be neglected, and patterns are selected by the local properties of
the wave equation. This is the regime of pattern formation, where 
the real parts of the interaction terms merely 
renormalize $h$ and $\omega_k$. The 
dominant limiting mechanism, in this case, is nonlinear damping\cite{Milner91}.
In  small systems, contrarily, the wave patterns are globally
determined by the boundary, and missphasing turns out to be more effective.

A system can be considered as small provided
that the dissipation time, $1/ \nu k^2$,
exceeds the typical time it takes a 
surface wave to cross the system. For capillary 
waves ($k> \sqrt{g \rho/\Sigma}$) 
this requirement yields the inequality
\begin{eqnarray}
\omega < \omega_c=\frac{ \Sigma}{ L \rho\nu}. \label{ssc}
\end{eqnarray}
In what follows we assume that this condition is satisfied, 
the wave patterns are determined by the boundary,
and missphasing is the dominant limiting mechanism\cite{comment}.

Consider now the Faraday waves in a container 
of non-integrable boundary shape having no discrete symmetries. 
The modes of the corresponding linear wave equation, 
$\psi_\alpha ({\bf r})$, are
eigenstates of a billiard system of the shape of the 
container. Thus, $\psi_\alpha ({\bf r})$ are approximately random 
Gaussian functions scarred by the periodic orbits of the underlying 
classical dynamics\cite{Heller84}. In principle, one can 
use $\psi_\alpha ({\bf r})$ as a basis for
the many body approach described above. However,
this basis is inconvenient. In the experiment, several modes are 
excited simultaneously (approximately 10), and 
the resulting wave patterns are determined by properties of the
short time dynamics of the system. Therefore, it 
is more natural to use scars as the skeleton of the 
excited wave patterns.

Below we develop a ``scar representation''. 
Instead of dealing with the amplitudes, $n_\alpha= |a_\alpha|^2$, 
of the eigenfunctions, $\psi_\alpha ({\bf r})$, we consider 
the amplitudes of scars. Using Wigner representation,
one can present ${\cal H}_0$ as
\begin{eqnarray}
 {\cal H}_0= \sum_\alpha \omega_\alpha n_\alpha =
\int \! \epsilon~ d \epsilon  \int \! d {\bf x}   W({\bf x}; \epsilon) 
~n({\bf x}) \label{Whamiltonian}
\end{eqnarray}
where ${\bf x}= ({\bf r}, {\bf k})$  is a phase space coordinate,
$n({\bf x})$  is the wave amplitude at ${\bf x}$,
and $W({\bf x}; \epsilon)$ is the Wigner representation of the 
imaginary part of the Green function\cite{PreviousTheory}:
\begin{eqnarray}
 W({\bf x}; \epsilon) = \sum_\alpha \delta (\epsilon \!-\!\omega_\alpha)
\!\int\! d {\bf q} ~e^{i {\bf k} \cdot {\bf q}} 
\psi_\alpha ({\bf r} \!-\! \frac{ \bf q}{2}) 
\psi_\alpha^* ({\bf r} \!+\! \frac{ \bf q}{2}). \nonumber 
\end{eqnarray}
Assuming the semiclassical limit $kL \gg 1$ as well as the 
capillary regime, $k> (g \rho/\Sigma)^{1/2}$,  we can approximate
$W({\bf x}; \epsilon)$ as
\begin{eqnarray} 
W({\bf x}; \epsilon) \simeq \delta [\epsilon - H_0({\bf x})]\left( 1+
\sum_p W_{p}({\bf x}; \epsilon) \right), \label{Wgreen}
\end{eqnarray}
where $H_0({\bf x}) = (\Sigma k^3 /\rho)^{1/2}$.
The first term in the parenthesis is the average of the random 
excitation over the energy shell.
The second term is a sum over the primitive periodic orbits of the system, 
$p$.  Each term of this sum represents the corresponding scar contribution:
\begin{eqnarray}
W_{p}({\bf x}; \epsilon)  \simeq 4 \pi \mbox{Re} 
\frac{ e^{ -\frac{u_p}{2} + i
( S_p \!-\! \varphi_p)}}{ 1 - e^{ -\frac{u_p}{2} + i
( S_p \!-\! \varphi_p)}} \tilde{\delta} ({\bf X}_p). \label{scarcontrib}
\end{eqnarray}
Here 
$S_p(\omega) \!=\! l_p (\rho \omega^2/\Sigma )^{1/3}$ is the action
of the periodic orbit with length $l_p$, instability 
exponent $u_p$, and  Maslov phase $\varphi_p$.
${\bf X}_p$ denotes a
coordinate on the  Poincare section of the periodic orbit. The function
$\tilde{\delta} ({\bf X}_p)$ is localized
along the $p$-th periodic orbit with the width of order 
$\sqrt{L/k}$. Its integral is unity so that  $\tilde{\delta} ({\bf X}_p)$
becomes a $\delta$-function in the classical limit\cite{PreviousTheory}
(i.e. at $k \to \infty$).

Substituting (\ref{Wgreen}) in (\ref{Whamiltonian}), we obtain
\begin{eqnarray}
{\cal H}_0= \int \! d \epsilon~\epsilon \bar{d}(\epsilon) \bar{n}(\epsilon)
+ \sum_{p,m} \omega_p^{(m)} n_p, \label{sch}
\end{eqnarray}
where $\bar{d}(\epsilon)= {\cal A}(\rho^2 \epsilon/ \Sigma^2)^{\frac{1}{3}}
/3 \pi$
is the mean density of states in a container of area 
${\cal A}$,
\begin{eqnarray}
\bar{n}= \frac{1}{\bar{d}} \int d {\bf x} \delta [\epsilon - H_0({\bf x})] 
n({\bf x})
\end{eqnarray}
is the average amplitude of the random background, and
\begin{eqnarray}
n_p = \frac{1}{t_p} \oint dt~ n({\bf x}_p(t)) 
\end{eqnarray}
is the amplitude of the scar $p$. Here $t_p$ is the period of 
the corresponding orbit, and ${\bf x}_p(t)$ is its coordinates 
parameterized by the time. The eigenfrequencies of the scars, 
$\omega_p^{(m)}$, are the poles of (\ref{scarcontrib}), i.e.~the
solutions of the equation $S_p(\omega_p^{(m)}) = 2 \pi  m + \varphi_p -iu_p/2$.
Expanding to leading order in $u_p$ we have  
\begin{eqnarray}
\omega_p^{(m)} \simeq \left(\frac{2 \pi m + 
\varphi_p}{l_p (\rho/\Sigma)^{1/3}}\right)^{3/2} - i \frac{\lambda_p}{2},
\label{wpm}
\end{eqnarray}
where $\lambda_p=u_p/t_p$ is the Lyapunov exponent 
of the orbit.

Now one can substitute the Hamiltonian (\ref{sch}) into the equations of 
motion (\ref{eqm}) and see that the Lyapunov exponent plays the role of a
dissipation. Indeed, $\lambda_p/2$ is the rate at which the ``particle 
escapes from the scar''. This rate should be added to the dissipation rate.
The Lyapunov exponents of few short periodic orbits of the stadium
billiard are tabulated in Table I. Comparison of these values with
the experimental bulk dissipation rate, $\gamma_b \simeq 2$ 
sec$^{-1}$, shows that their effect is significant. 

Next, we consider the dissipation 
at the side walls of the container\cite{Benjamin54}. 
This dissipation has been calculated for particular cases, e.g., 
the square and the circular containers\cite{boundarydis}.  From these 
results it is evident that boundary dissipation 
depends on the particular form of the wave pattern.
To obtain the result for a general container shape, and 
arbitrary scarring pattern, we first note that boundary dissipation
takes place very close to the boundary, at distance of order 
$l_D= \sqrt{2 \nu/\omega} \sim 10^{-2}$cm. Viewing the scar 
as a plain wave scattered from the boundary, each collision point
adds its own contribution to the dissipation rate.
These contributions can be calculated 
by traditional methods\cite{boundarydis}. Assuming
pinning of the fluid surface at the boundary, the resulting
dissipation rate of a scar can be presented as  a sum 
over the reflection points of the orbit from the 
boundary:
\begin{eqnarray}
\gamma_p \simeq \frac{( 2 \nu \omega)^{1/2} } {l_p}
\sum_n \frac{ 1 - \frac{1}{2} \cos^2(\theta_n)}{\cos (\theta_n)}, \label{bd}
\end{eqnarray}
where $\theta_n$ is the angle between the orbit and 
the normal to the boundary. Examples of $\gamma_p$ for various scars
of the stadium billiard are presented in Table I. Note that the 
anomalously large dissipation rate of whispering gallery modes
(patterns Nos. 8, 9) results from $\theta_n$ being close to
$\pi/2$, see (\ref{bd}).

The boundary dissipation of the random background, $\bar{\gamma}$,
can be obtained from the result for a long ergodic orbit. 
However, the result diverges since  ergodicity implies
that $\theta_n$ can become arbitrarily close to $\pi/2$.
Cutting this divergence by the wave resolution, 
$\Delta \theta_n \approx 1/kL$, we obtain
\begin{eqnarray}
\bar{\gamma} \simeq \frac{( 2 \omega \nu)^{1/2}}{L} \ln ( k L).
\end{eqnarray} 
{\narrowtext   
\begin{table} [ht]     
\begin{center}     
\leavevmode       
        \epsfxsize=8.5cm      
         \epsfbox{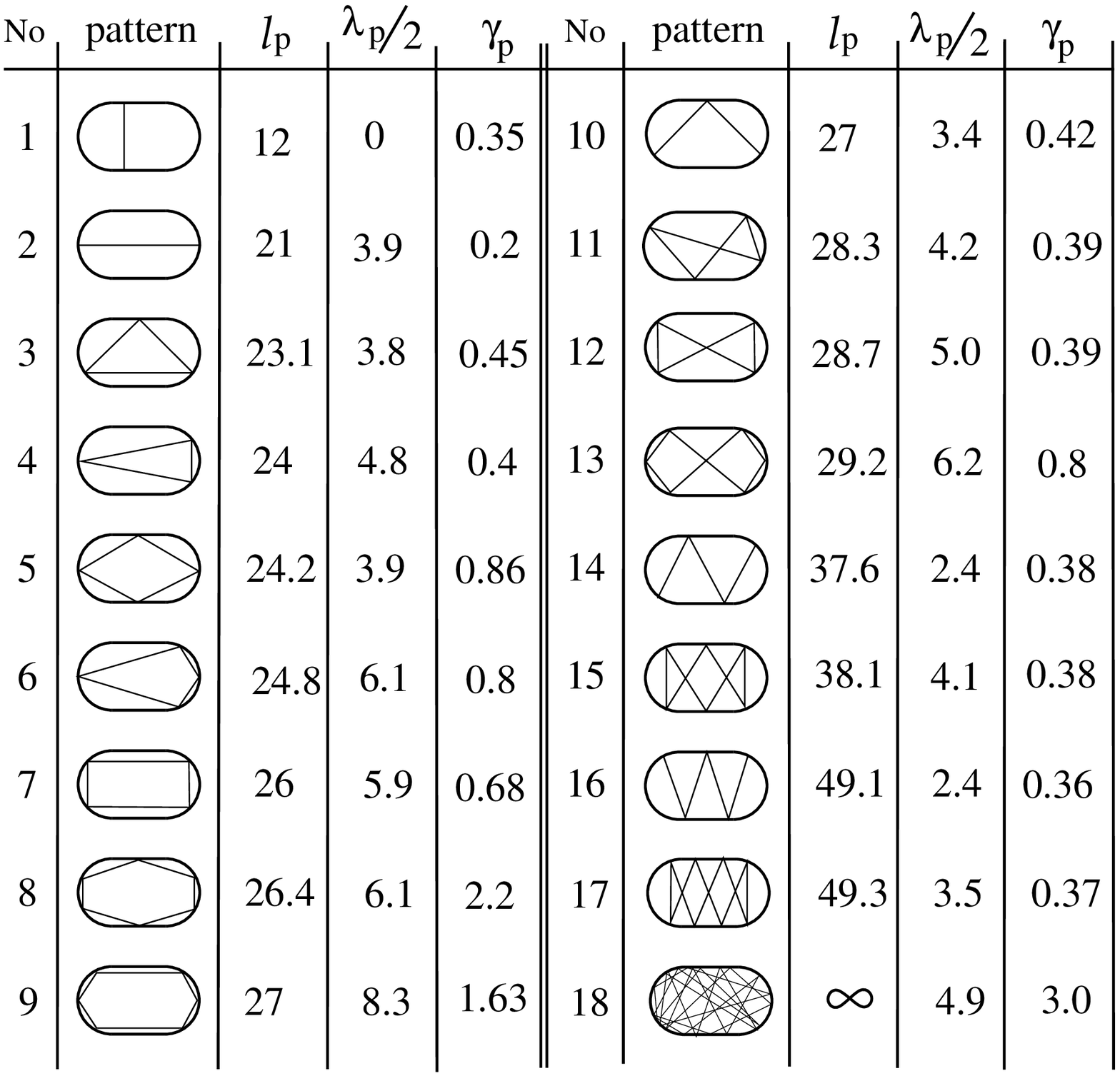}      
\end{center}    
	\caption{Parameters of some periodic orbits of the 
stadium billiard, with  dimensions of the experimental system:
semicircles radius 3 cm, and distance between semicircles 4.5 cm.
$l_p$ is the length of the orbit in centimeters, $\lambda_p$ is the
Lyapunov exponent in sec$^{-1}$, and $\gamma_p$ is the contribution to the 
dissipation rate due to the boundary in sec$^{-1}$. 
The velocity is set to be similar to the experimental value, 66 cm/sec.}   
\end{table}   
} 
\noindent

Collecting the various dissipation terms and the contribution 
of the Lyapunov exponent, the threshold for excitation of scars 
at resonance $(\omega=\mbox{Re}~ \omega_p^{(m)})$ is:  
\begin{eqnarray}
 h>h_p= \gamma_b+ \gamma_p+\lambda_p/2
\end{eqnarray}
Therefore, large boundary dissipation 
prevents the excitation of random Gaussian patterns,
as well as other modes such as the whispering gallery modes. 
Moreover, increasing the 
boundary dissipation, e.g. by 
lowering the fluid level, should 
eventually suppress all of the scars except 
the horizontal one (No.~2 in Table I). This is 
precisely what is observed in the experiment\cite{exp}. 

Turning to the interaction effects, we first note that
matrix elements
off diagonal in scars, $I_{p,p'}^{(j)}$,
are usually much smaller than the diagonal ones, 
$I_{p,p}^{(j)}$. The reason is that scars are localized objects, 
whereas the interaction between any two of them
is proportional to their intersection area. 

Consider, therefore, the scattering from a scar to itself. Here, the
wave vectors, $k$ and $k'$, are either parallel or antiparallel,
except for small regions of self intersections. Thus the
diagonal matrix elements of a scar associated with an orbit, $p$,
which has a time reversal counterpart $-p$ (such as 
Nos.~3--9 in Table I) are $I^{(1)}_{p,p} \simeq I_k^{(1)}(0)$, 
$I^{(1)}_{p,-p}\simeq
I_k^{(1)}(\pi)$ and $I^{(2)}_{p,p}\simeq I_k^{(2)}(0)$. 
The interaction matrix elements of scars associated with self 
retracing orbits (such as No.~1, 2, and 10 in Table I) are 
$I_{p,p}^{(1)}=I_{p,p}^{(2)}\simeq (I^{(1)}_k(0)+  
I^{(1)}_k(\pi)+ I^{(2)}_k(0))/2$.

If only one scar is excited, its amplitude can be evaluated
exactly from the steady state limit of Eq.~(\ref{eqm}).
For scars associated with self retracing orbits, the result is
\begin{eqnarray}
n_p^{(1)}=  \frac{1}{2 T}\sqrt{ h^2- h_p^2},
\end{eqnarray} 
where $T=\mbox{Re}~I_{p,p}^{(1)}$. We assumed that the system 
was at resonance, 
$\omega\!=\!\mbox{Re}~\omega_p$, and neglected
the imaginary part of $I_{p,p}^{(j)}$.

Let now two scars be in resonance with the driving frequency, 
one can calculate the amplitude of
the scar $p$ in the presence of the second scar $p'$ by perturbation 
theory in the off diagonal matrix element between these scars. 
The result is
\begin{eqnarray}
n_p^{(2)}\simeq n_p^{(1)} \left(1- 
\tau_1 Q- \tau_2 Q^2 \right) \label{2scars}
\end{eqnarray}
where $\tau_j= \mbox{Re} I_{p p'}^{(j)}/2T$ and
$Q=n_{p'}^{(1)}/n_{p}^{(1)}$. According to
Eq.~(\ref{2scars}) even a 
relatively weak interaction
can suppress the scar with the higher threshold amplitude.
Thus one should expect to observe only the dominating scar.

Finally, we comment on the role of discrete symmetries in the shape of
the container. The exact eigenfunctions in this case can
be classified according to the irreducible representation of the 
corresponding symmetry group. 
In the case of the stadium billiard this symmetry is not 
sufficient to produce an exact degeneracy of eigenvalues. However, the 
eigenfrequencies of scars, associated with an excitation 
of a set of modes, are degenerate. The scars associated with 
orbits Nos.~3, 4 and 11 in Table I can be used as examples.
The symmetry between the scars is
usually weakly broken by perturbations such as 
inhomogeneous pumping or deviation from an exact horizontal 
state of the container. Interaction, in turn, 
drives the system into a nonsymmetric wave pattern.

To summarize, we have shown that Faraday wave patterns in a container of
non-integrable boundary shape are mainly scars. The instability threshold 
for these patterns is determined by  bulk dissipation,  Lyapunov 
exponents, and dissipation near the container side walls. 
The latter is too strong for random patterns 
and whispering gallery modes to appear.
Nonsymmetric wave patterns, in a symmetric container, take place
due to the existence of degenerate scars
and interaction effects between them.

We are grateful to Arshad Kudrolli and Jerry Gollub
for many discussions and communications, and for 
the experimental data presented in this Letter. 
This research was supported by Grant No.~9800065 from 
the USA-Israel Binational Science Foundation (BSF). 

\end{multicols}

\end{document}